\documentclass[12pt,preprint]{elsarticle}
\usepackage{epsfig,amssymb,bm,graphics,color,amsmath}
\usepackage{graphicx}
\usepackage{slashed}
\usepackage{chngcntr}

\counterwithout{footnote}{section}

\DeclareMathOperator{\erf}{erf}

\begin{document}{\normalsize}

\title{
Equation of State and Viscosities
from a Gravity Dual of the Gluon Plasma}
\author{R. Yaresko and B. K\"ampfer}
\address{Helmholtz-Zentrum Dresden-Rossendorf,
POB 51 01 19, 01314 Dresden, Germany\\
TU Dresden, Institut f\"ur Theoretische Physik, 01062 Dresden, Germany}


\begin{abstract}
Employing new precision data of the equation of state 
of the SU(3) Yang-Mills theory (gluon plasma) 
the dilaton potential of a gravity-dual model
is adjusted in the temperature range $(1 - 10) T_c$
within a bottom-up approach. 
The ratio of bulk viscosity to shear viscosity follows then
as $\zeta/\eta \approx \pi \Delta v_s^2$ for $\Delta v_s^2 < 0.2$
and achieves a maximum value of $0.94$ at $\Delta v_s^2 \approx 0.3$, 
where $\Delta v_s^2 \equiv 1/3 - v_s^2$ is the
non-conformality measure and $v_s^2$ is the velocity of sound squared, while the ratio of shear
viscosity to entropy density is known as $(4 \pi)^{-1}$
for the considered set-up with Hilbert action
on the gravity side.
\end{abstract}

\maketitle

\section{Introduction}
With the advent of new precision data \cite{WuppertalBp}, which extend previous
lattice QCD gauge theory evaluations \cite{Boyd,Okamoto} for the pure gluon plasma to a larger
temperature range, a tempting task is to seek for an appropriate gravity dual model. 
While such an approach does not neccessarily provide new insights in the
pure SU(3) Yang-Mills equation of state above the deconfinement temperature
$T_c$, it however allows to calculate, without additional ingredients, further
observables, e.g.\ transport coefficients. 
(This is in contrast to quasiparticle approaches which require
additional input to access transport coeffcients \cite{Bluhm}.) 
In considering an ansatz of gravity+scalar as framework of effective 
dual models to pure non-abelian gauge thermo-field theories
within a bottom-up approach one has to adjust 
either the potential of the dilaton field, 
or a metric function, or the dilaton profile. 

The improved holographic QCD (IHQCD) model, 
developed in \cite{Kiritsis_first1, Kiritsis_first2, Kiritsis_long, Kiritsis_compwithdata} 
(for a review cf.\ \cite{Kiritsis}) 
is a particularly successful realisation of such a setting. 
The potential of IHQCD \cite{Kiritsis_compwithdata} was constructed to 
match the t'Hooft limit Yang-Mills $\beta$ function to two-loop order (which determines the functional 
form and two parameters) in the near-conformal (small t'Hooft coupling) region, while
the zero-temperature (large t'Hooft coupling) behavior is fixed by 
demanding confinement and a linear glueball spectrum. 
A potential smoothly interpolating between the two asymptotic regions was shown in 
\cite{Kiritsis_compwithdata} to well reproduce the $N_c = 3$ Yang-Mills plasma 
equation of state \cite{Boyd},  
where remaining free parameters were fixed by comparing 
to the latent heat and scaled pressure from the lattice. 
Within IHQCD, zero-temperature confining geometries exhibit a first-order thermodynamic 
phase transition \cite{Kiritsis_long}.

A different type of dilaton potentials was considered in \cite{Gubser}, where near the boundary 
the potential accounts for a massive scalar field and the spacetime asymptotes to pure $AdS_5$. 
The potential parameters were matched to the velocity of sound as suggested by the 
hadron resonance gas model and the dimension of $\text{Tr} \, F^2$ at a finite scale \cite{Gubser_PRL} 
and reproduce the velocity of sound of 2+1 flavor QCD, 
whereas in \cite{Noronha} the matching to the SU(3) Yang-Mills equation of state \cite{Boyd} 
has been accomplished. 
In IHQCD, the marginal operator dual to $\phi$ is $\text{Tr}\, F^2$ \cite{Kiritsis_first1}, 
while in \cite{Gubser,Gubser_PRL} and here the dual operator $\mathcal{O}$ is interpreted as 
a relevant deformation of the boundary theory Lagrangian. 
In \cite{Mei}, instead of the dilaton potential, an ansatz for a metric function
of the five-dimensional gravity action is selected and consequences for the 
boundary theory are explored (Such an approach suffers however from the conceptual 
shortcoming that the dilaton potential and thus the action depend on the temperature, 
while, according to the gauge/gravity duality, the bulk action should be independent of the boundary theory state).

The previous benchmark lattice data \cite{Boyd} (up to $4.5 T_c$)
and further SU($N_c$) data for $N_c \le 8$ \cite{Panero} (up to $3.5T_c$) 
and $N_c \le 6$ \cite{DattaGupta} (up to $4T_c$) 
are for $N_c = 3$ now supplemented and extended up to $1000T_c$ \cite{WuppertalBp}.
Here we are going to adjust precisely the dilaton potential 
to the new lattice data \cite{WuppertalBp} in the temperature range up to $10T_c$, 
thus catching the strong-coupling regime, as envisaged as relevant also in \cite{Kajantie}. 
We discard completely a recourse to the $\beta$ function. 
Such an approach can be considered as a
convenient parameterization of the equation of state. Once the potential is adjusted, 
it qualifies for further studies, e.g.\ of transport coefficients. Our goal is accordingly 
the quantification of the bulk viscosity in the
LHC relevant region, in particular near to $T_c^+$, and a comparison with results
of the quasiparticle model \cite{Bluhm}. \pagebreak

According to holography, $SU(N_c)$ Yang-Mills theory at finite $N_c$ must be 
described by quantum string theory, which has not yet been completely established. 
Since in the large-$N_c$ and large t'Hooft coupling limits quantum  
string theory reduces to classical gravity, one presently resorts to a gravitation 
theory in a five-dimensional space, constructed in such a manner 
to accomodate certain selected features of the holographically 
emerging boundary field theory. 
As in the models \cite{Kiritsis, Gubser}, one often considers the AdS/QCD correspondence 
as deformation of the original AdS/CFT correspondence \cite{Maldacena_original} by additional (relevant or marginal) 
operators which allow a qualitative study of QCD or Yang-Mills properties in the strong-coupling 
regime. 
Conclusions for the latter theories should be drawn with caution:  
For instance, in the perturbative regime 
of the (large-$N_c$) boundary theory, the gravity theory is expected to become strongly coupled 
and, consequently, finite string scale corrections may arise; if one also leaves 
the t'Hooft limit, stringy loop corrections may matter.
(It is known that equilibrium thermodynamics of $SU(N_c)$ Yang-Mills depend only 
weakly on $N_c$, see \cite{Panero} and references therein.) 
Having these disclaimers in mind, we nevertheless study quantitatively 
the bulk viscosity in a bottom-up setting matched solely to $N_c = 3$ Yang-Mills 
thermodynamics within $(1-10)T_c$. 

The potential asymptotics of Gubser-Nellore \cite{Gubser,Gubser_PRL} and IHQCD \cite{Kiritsis} 
models are different 
both in the near-boundary region i.e.\ at high temperatures 
and also deep in the bulk i.e.\ at low temperatures. When 
adjusting the potential in an intermediate region suitable for $(1-10)T_c$ one would like to 
know whether it is important to incorporate a certain kind of asymptotics, or whether 
they have little influence. Put another way, to what extent does a fit to lattice 
data on $(1-10)T_c$ determine the potential? Here, we do not attempt to solve the 
general problem of computing the potential from a given equation of state, but 
instead show that various potentials which 
contain a certain unique relevant section
lead to nearly identical equations of state in the corresponding temperature region. 

Transport properties of the matter produced in relativistic heavy-ion collisions at
RHIC and LHC are important to characterize precisely such novel states of a
strongly interacting medium besides the equation of state.
The impact of the bulk viscosity on the particle spectra and differential elliptic
flows has been recently discussed in \cite{DS} and found to be sizeable 
in \cite{DS_prime}, in particular for higher-order collective flow harmonics.
The bulk viscosity enters also
a new soft-photon emission mechanism \cite{Dima} via the conformal anomaly,
thus offering a solution to the photon-$v_2$ puzzle
(cf.\ \cite{Dima} for details and references).
Compilations of presently
available lattice QCD results of viscosities can be found in \cite{Bluhm}. 

\section{The set-up}
\label{sec:setup}
The action 
$S = \frac{1}{16 \pi G_5} \int d^5 x \sqrt{-g}
\left\{R - \frac12 (\partial \phi)^2 - V(\phi) \right\}$
(the Hawking-Gibbons term is omitted) leads, with the ansatz
for the infinitesimal line element squared in Riemann space
$ds^2 = \exp\{2 A\} (d\vec x^2 - h dt^2) + \exp\{2 B\} h^{-1} L^2 d\phi^2$, 
to the field equations quoted in \cite{Gubser} under (25a - 25c); the equation
of motion (25d) follows from the derivative of (25c) with insertion of
(25a - 25c). Here, the coordinate transformation 
$dz = L  \exp\{ B - A \} d \phi$ 
has been employed to
go from the Fefferman-Graham coordinate $z$ 
in the infinitesimal line element squared
$ds^2 = \exp\{2 A\} ( - h \,dt^2 + d\vec x^2 + h^{-1} \,dz^2)$
to a gauged radial
coordinate expressed by the dilaton field $\phi$ which requires 
the introduction of a length scale $L$.
The metric functions are thus to be understood as 
$A(\phi; \phi_H)$, $B(\phi; \phi_H)$ and $h(\phi; \phi_H)$, and a prime means in the following the
derivative with respect to $\phi$. These equations can be rearranged 
by defining $Y_1 = A - A_H$, $Y_2 = A^\prime + U$, 
$Y_3 = A^{\prime \prime} + \frac{1}{2} U^\prime$, $Y_4 = B - B_H$, 
$Y_5 = \exp(4A_H-B_H) \int_{\phi_H}^\phi d\tilde{\phi}\exp(-4A+B)$, 
where the subscript $H$ denotes the value of a function at the horizon and $U \equiv V / (3 V')$,
to change the mixed boundary value problem into an initial value problem, given by
\begin{eqnarray}
Y_1'&=&Y_2 - U, \label{eq:Y_1'} \\
Y_2' &=& Y_3 + \frac12 U', \label{eq:Y_2'}\\
Y_3' &=& \frac12 U'' + \frac{Y_3- \frac12 U'}{(Y_2 -U)Y_2} \label{eq:Y_3'}
\left( (Y_3 - \frac12 U') (3 Y_2 - 2 U) +
(4Y_2 - \frac{U'}{U})(Y_2 - U)^2
\right. \nonumber\\
&& \hspace*{4cm} \left. +
\frac{Y_2}{6 U} (2 Y_2 - U)
\right),\\
Y_4' &=& \frac{ 6 (Y_3 - \frac12 U') + 1}{6 (Y_2 - U)}, \label{eq:Y_4'} \\
Y_5' &=& \exp\{-4 Y_1+Y_4\} \label{eq:Y_5'}
\end{eqnarray}
which is integrated from the horizon $\phi_H-\epsilon$, to the boundary $\epsilon$ 
with the initial values $Y_i = 0$
at $\phi_H-\epsilon$. The limit $\epsilon \to 0^+$ has to be taken to obtain
the entropy density $s$ and the temperature $T$  
\begin{eqnarray}
G_5 s &=& \frac14 \exp(3 A_H), \\
L T &=& - \frac{1}{4 \pi} \frac{\exp(A_H-B_H)}{Y_5(\epsilon)}, \label{eq:T_H}
\end{eqnarray}
where $A_H = \frac{\log \epsilon}{\Delta-4}- Y_1(\epsilon)$ and
$B_H = -\log(-\epsilon [\Delta-4]) - Y_4 (\epsilon)$.
This set\footnote[1]{The system (\ref{eq:Y_1'}-\ref{eq:Y_5'}) enjoys some redundancy. 
It can be reduced by introducing 
$X \equiv 1/4A^\prime = 1/(4[Y_2 - U])$, 
$Y \equiv h^\prime/(4 h A^\prime) = Y_5^\prime/(4Y_5[Y_2 -U])$ 
which leads to two coupled first-order ODEs for 
the scalar invariants $X(\phi; \phi_H)$ and $Y(\phi; \phi_H)$ according to 
\cite{Kiritsis_long}. Eliminating $Y$ in this system leads to a second-order 
ODE for $X$, equivalent to the ``master equation'' in \cite{Gubser}. 
Two additional quadratures are then needed to obtain the thermodynamics via 
$LT = \frac{V(\phi_H)}{\pi V(\phi_0)} 
\exp(A(\phi_0) + \int_{\phi_0}^{\phi_H} d\phi [ \frac{1}{4X} + \frac23 X])$ and 
$G_5s = \frac{1}{4} \exp(3A(\phi_0) + \frac34\int_{\phi_0}^{\phi_H} d\phi \frac{1}{X})$.
The set (\ref{eq:Y_1'}-\ref{eq:Y_5'}) does not need such additional quadratures.
}
ensures the boundary conditions
$h(\phi=0) = 1$ and $h(\phi_H) = 0$ as well as the AdS asymptotic limits
$A (\phi) =\frac{\log \phi}{\Delta-4}$ (we set $L\Lambda = 1$ \cite{Gubser}) and
$B (\phi) = -\log(-\phi [\Delta-4])$
at $\phi \to 0^+$. The boundary asymptotics of $A$ and $B$ assume 
$L^2V(\phi) \approx -12 + (\Delta [\Delta - 4]/2)\phi^2$ for small $\phi$,  
where $\Delta$ is the scaling dimension of the conformality-breaking operator of the 
boundary theory. We consider $2 < \Delta < 4$, selecting the upper branch of the 
mass dimension relation $L^2 M^2 = \Delta (\Delta - 4)$ and restricting to relevant operators. 
Hence, the Breitenlohner-Freedman bound $L^2 M^2 \ge -4$ \cite{BF_bound} is respected 
and renormalizability on the gauge theory side is ensured. 
The quantities $Y_i (\epsilon)$ depend on the horizon position $\phi_H$,
implying in particular $s(\phi_H)$ and $T(\phi_H)$,
thus providing the equation of state $s(T)$ in parametric form. 

\section{Equation of state} 
\label{sec:EoS}
To compare with the lattice results \cite{WuppertalBp} of 
the relevant thermodynamical quantities
(i) sound velocity squared $v_s^2 = \frac{d \log T}{d \log s}$,
(ii) scaled entropy density $s/T^3$,
(iii) scaled pressure $p/T^4$, and
(iv) scaled interaction measure $I/T^4 = s/T^3 - 4 p/T^4$
(all as functions of $T/T_c$)
one must adjust the scale $T_c$ and the 5D Newton`s constant
$G_5$ (actually, the dimensionless combinations
$LT_c$ and $G_5/L^3$ are needed).
In the present bottom-up approach, we employ a new potential 
designed to reproduce the data \cite{WuppertalBp} in the temperature 
region $(1 - 10)T_c$,
\begin{equation} \label{eq:v1}
 v_1(\phi) = \frac{V^\prime(\phi)}{V(\phi)}= 
 \begin{cases}
   \frac{- L^2 M^2}{12}\phi + i_1 \phi^3 &\text{ for } \phi \leq \phi_m,  \\
   \gamma + s_1[\erf(s_2( \phi - s_3)) - 1]  &\text{ for } \phi \geq \phi_m,
 \end{cases}
\end{equation}
as an ansatz and optimize the parameters 
$\phi_m$, $s_{1,2,3}$ and $\gamma$.
Since we are not interested in the high-temperature regime $T > 10T_c$, we choose 
a simple interpolation from $\phi = 0$ to $\phi = \phi_m$. The latter value 
is taken as a fit parameter and fixes $L^2 M^2$ and $i_1$ by the requirement 
that $v_D$ should be differentiable at $\phi_m$. 
The critical temperature $LT_c$
is determined by $T_c = T(\phi_H^c)$ with $\phi_H^c$ from the pressure 
\begin{equation} \label{eq:p}
 p(\phi_H) = \int_{\infty}^{\phi_H} d\tilde\phi_H s(\tilde\phi_H) \frac{dT}{d\tilde \phi_H},
\end{equation}
via $p(\phi_H^c) = 0$. 
This is the prescription discussed in detail in 
\cite{Kiritsis_long} for the first-order phase transition to a 
thermal gas configuration at $T < T_c$. According to \cite{Kiritsis_first2, Kiritsis_long} 
the boundary theory at $T < T_c$ is confining and gapped 
if $\gamma > \sqrt{2/3}$ and, equivalently, $LT(\phi_H)$ is 
U shaped, with a global minimum at $\phi_H^{min}$, implying $T(\phi_H^c) > T(\phi_H^{min})$, 
see Fig.~A.3. The construction ensures a minimum free energy for $T < T_c$ (thermal gas with $p = 0$) 
and $T > T_c$ (large black hole branch which continues in the UV region).
In \eqref{eq:p}, $p(\infty) = 0$ for a ``good'' IR singularity requires   
$\gamma < 2 \sqrt{2/3}$. 

\begin{figure}[!ht]
\begin{center}
\includegraphics[width=0.495\textwidth]{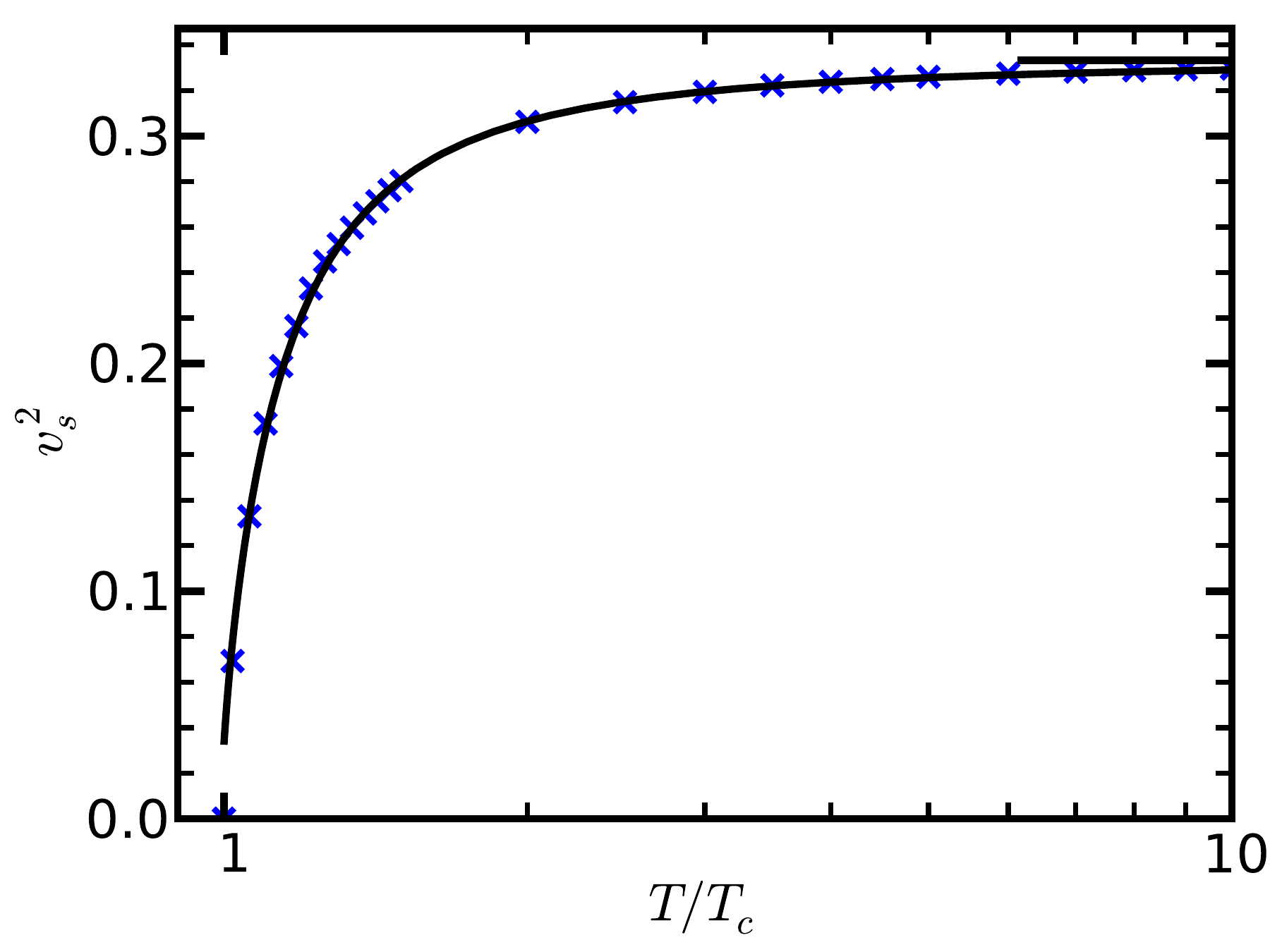}
\includegraphics[width=0.495\textwidth]{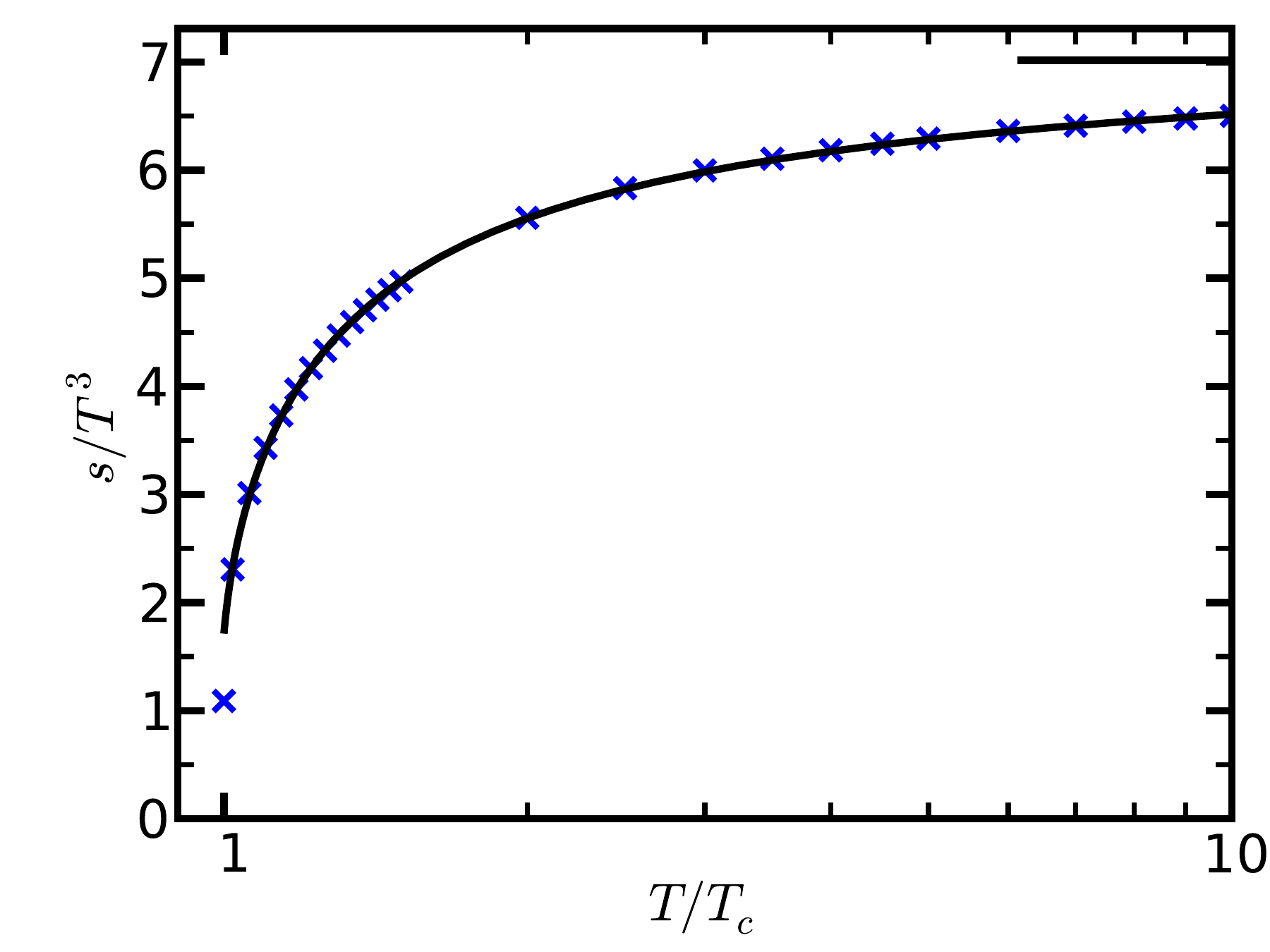}
\includegraphics[width=0.495\textwidth]{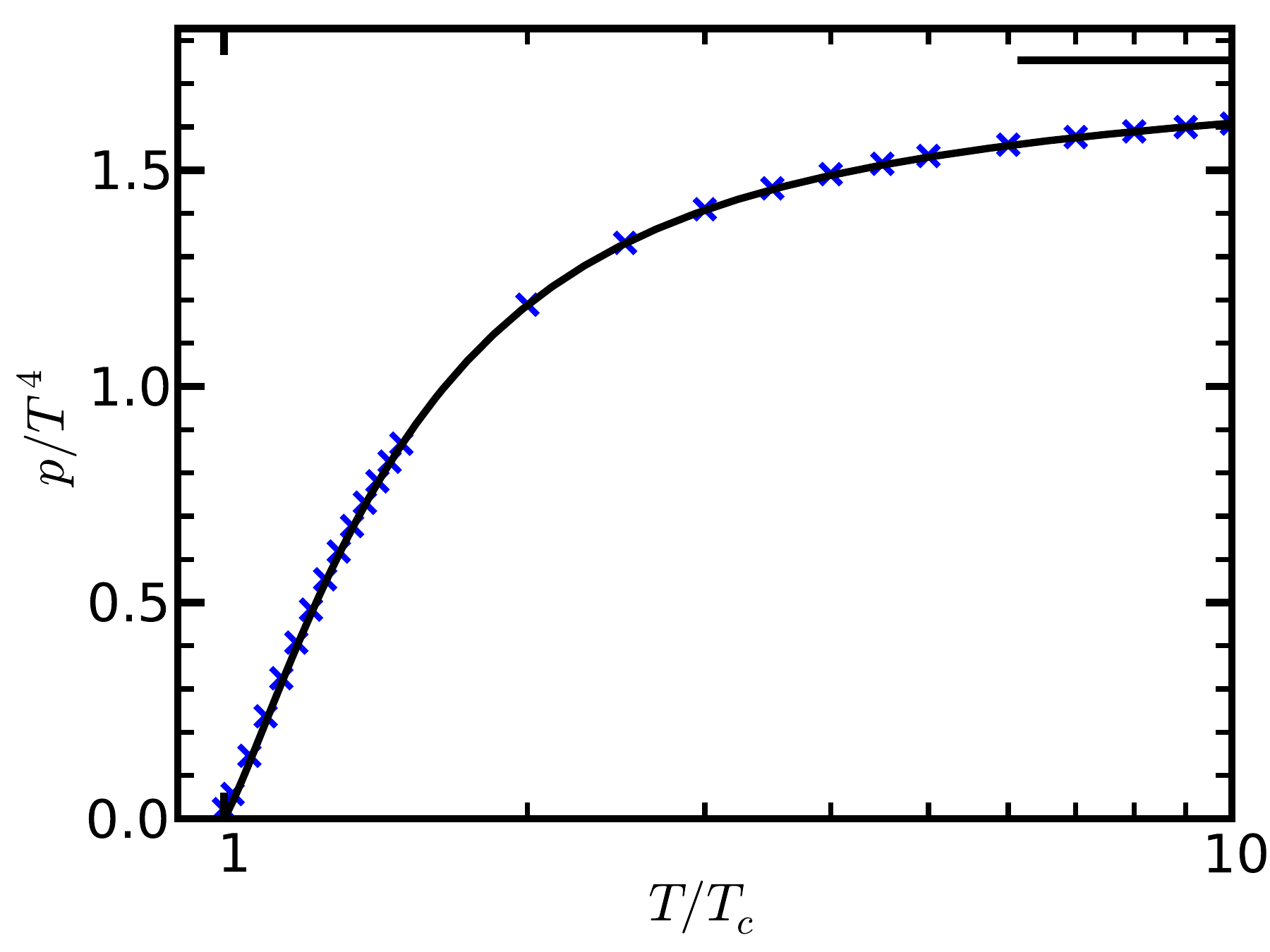}
\includegraphics[width=0.495\textwidth]{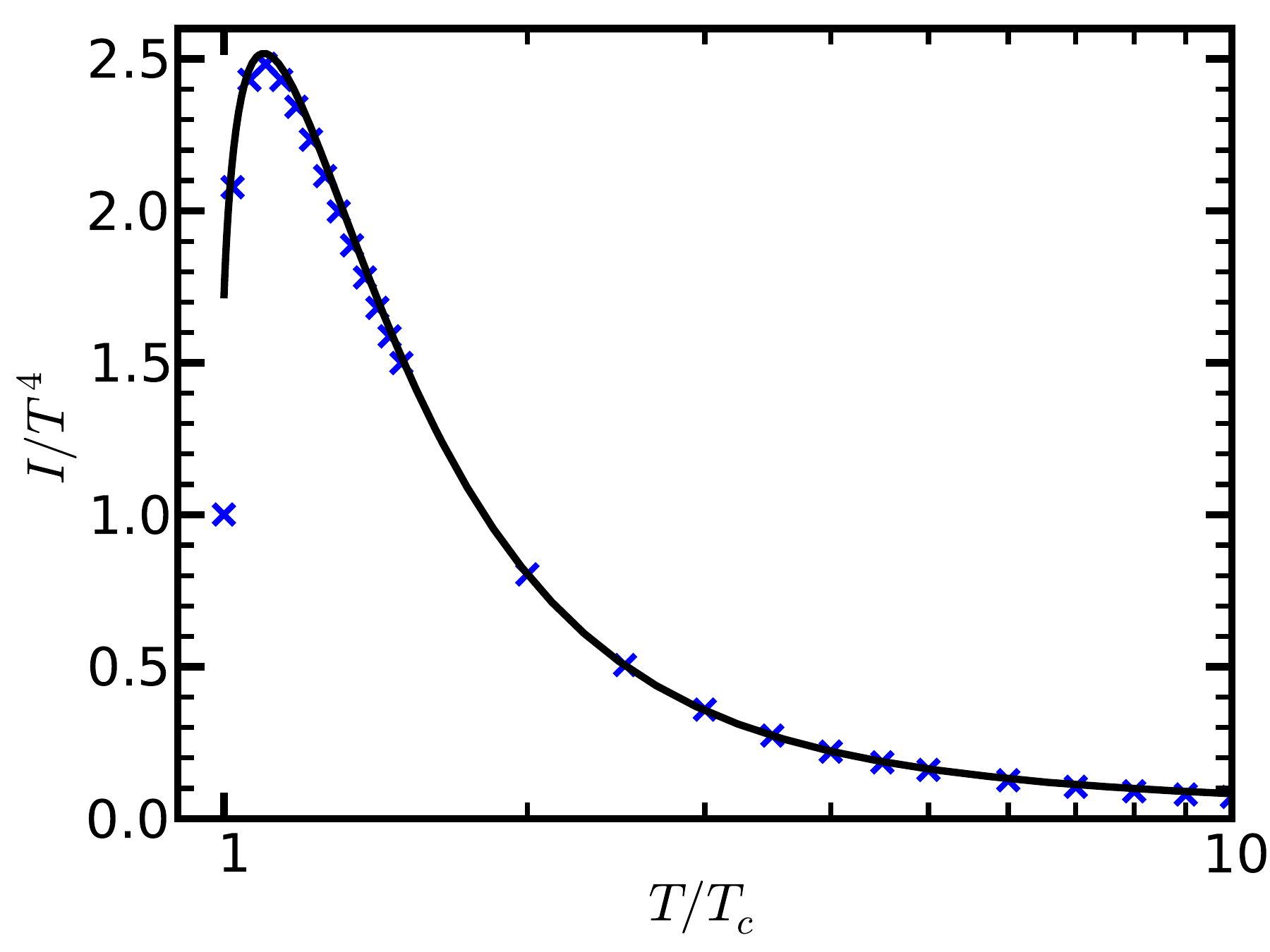}
\end{center}
\caption{
The sound velocity squared $v_s^2$ (left top panel),
scaled entropy density $s/T^3$ (right top panel),
scaled pressure $p/T^4$ (left bottom panel), and
scaled interaction measure $I/T^4$ (right bottom panel)
as functions of $T/T_c$
for the potential $v_1$ \eqref{eq:v1} with optimized parameters \eqref{eq:v1_pars}. 
The lattice data (symbols) are from \cite{WuppertalBp}.
The horizontal lines in the upper right corners depict the respective Stefan-Boltzmann limits.  
\label{fig.1}}
\end{figure}
 
Our results are exhibited in Fig.~1 for the optimized parameter set 
\begin{equation} \label{eq:v1_pars}
\begin{tabular}{c|c|c|c|c|c|c}
   $v$      & $\phi_{m}$ &$s_{1}$ &$s_{2}$ &$s_{3}$ & $\gamma$    & $G_5/L^3$ \\ \hline
  $v_1$     & 1.3444     & 0.3954 & 0.6723 & 2.7358 & 0.8222         & 1.1100 
\end{tabular} .
\end{equation}
The velocity of sound is independent of $G_5$ which steers the number of degrees
of freedom, thus being important for entropy density, energy density $e$, pressure
and interaction measure. In asymptotically free theories, the
$T^4$ term dominates $s$, $e$ and $p$ at large temperatures; it is subtracted in 
the interaction measure making it a sensible quantity. 
(Unlike the IHQCD model our ansatz does not catch pQCD features in the deep UV. 
That is the reason for our restriction to $T < 10 T_c$.)
The appearance of a maximum
of $I/T^4$ at $T/T_c \approx 1.1$ 
is related to a turning point of $p/T^4$ as 
a function of $\log T$. Position and height of $I/T^4$ -- the primary quantity
in lattice calculations -- are sensible characteristics of the equation of state.
The dropping of $I/T^4$ at larger temperatures signals the approach towards 
conformality. (Since in conformal theories $v_s^2 = 1/3$, the quantity 
$\Delta v_s^2 = 1/3 - v_s^2$ is termed non-conformality measure; also here, the
dominating $T^4$ terms at large temperatures drop out.)
Inspection of Fig.~1 unravels the nearly perfect description of the lattice data
\cite{WuppertalBp}. Note that, by construction, $p/T^4$ 
always slightly underestimates the lattice data for $T \to T_c^+$, 
since $p(\phi_H^c) = 0$, while $p(T_c)/T_c^4\vert_{lattice} = 0.0222$ \cite{WuppertalBp}.
We find $\Delta s(T_c)/T_c^3 \approx 1.7$ for the scaled latent heat.

\section{Viscosities} 
Irrespectively of the dilaton potential $V(\phi)$, 
the present set-up with Hilbert action $R$ for the
gravity part delivers $\eta / s = (4 \pi)^{-1}$ \cite{Gubser_visc, EO}
for the shear viscosity $\eta$, often denoted as KSS value \cite{KSS}. 
(See \cite{KSS_orig} for the original calculation. 
Inclusion of higher-order curvature corrections can decrease the KSS 
value \cite{Buchel_eta_violation}.) 
In contrast, 
the bulk viscosity to entropy density ratio $\zeta/s$ 
has a pronounced temperature dependence. 
Following \cite{Gubser_visc} we calculate $\zeta$ from the relation
\begin{equation}
\frac{\zeta}{\eta}\Big\vert_{\phi_H} = \frac{1}{9 U(\phi_H)^2}
\frac{1}{\vert p_{11}(\epsilon) \vert^2} , \label{eq:zeta_GPR}
\end{equation}
where the asymptotic value $p_{11}(\epsilon)$ of the perturbation $p_{11}$ of the 
11-metric coefficient is obtained by integrating 
\begin{equation}
p_{11}'' + 
\left( \frac{1}{3 (Y_2 - U)} + 4(Y_2 - U) - 3 Y_4' + \frac{Y_5'}{Y_5} \right) p_{11}' +
\frac{Y_5'}{Y_5} \frac{Y_3 - \frac12 U'}{Y_2 - U} p_{11} = 0
\end{equation}
from the horizon $\phi_H - \epsilon$ to the boundary $\epsilon$ with initial conditions
$p_{11} (\phi_H - \epsilon) = 1$ and $p_{11}' (\phi_H - \epsilon) = 0$
and $\epsilon \to 0^+$. Equivalently \cite{Buchel_Kiritsis}, the bulk viscosity 
can be obtained from the Eling-Oz formula \cite{EO}
\begin{equation}
 \frac{\zeta}{\eta}\Big\vert_{\phi_H} = \Big( \frac{d \log s}{d \phi_H} \Big)^{-2} = \Big( \frac{1}{v_s^2}\frac{d \log T}{d \phi_H} \Big)^{-2} . \label{eq:zeta_EO}
\end{equation}
Our results are exhibited in Fig.~2. The scaled bulk viscosity $\zeta /T^3$
has a maximum at $1.05 T_c$ 
(which is slightly below the maximum of $I/T^4$) 
and drops rapidly for increasing temperatures, see left panel of Fig.~2. 
Remarkable is the almost linear section
of $\zeta / \eta$ as a function of the non-conformality measure $\Delta v_s^2$
(see right panel), as already 
suggested in \cite{Buchel} and observed, in particular at high temperatures, 
in numerous holographic models \cite{Buchel_bound, Cherman_Yarom};
for further reasoning on such a linear behavior within holography approaches
cf.\ \cite{Skenderis}. 
A non-linear behavior occurs in a small temperature
interval $1 \le T/T_c < 1.05$, i.e.\ for  $\Delta v_s^2 > 0.22$,
see right panel of Fig.~2.
The maximum value of $\zeta / \eta \approx 0.94$ at $\Delta v_s^2 \approx 0.3$ 
depends fairly sensitively on the details of the equation of state for $T \to T_c^+$. 

\begin{figure}[!h]
\begin{center} 
\includegraphics[width=0.495\textwidth]{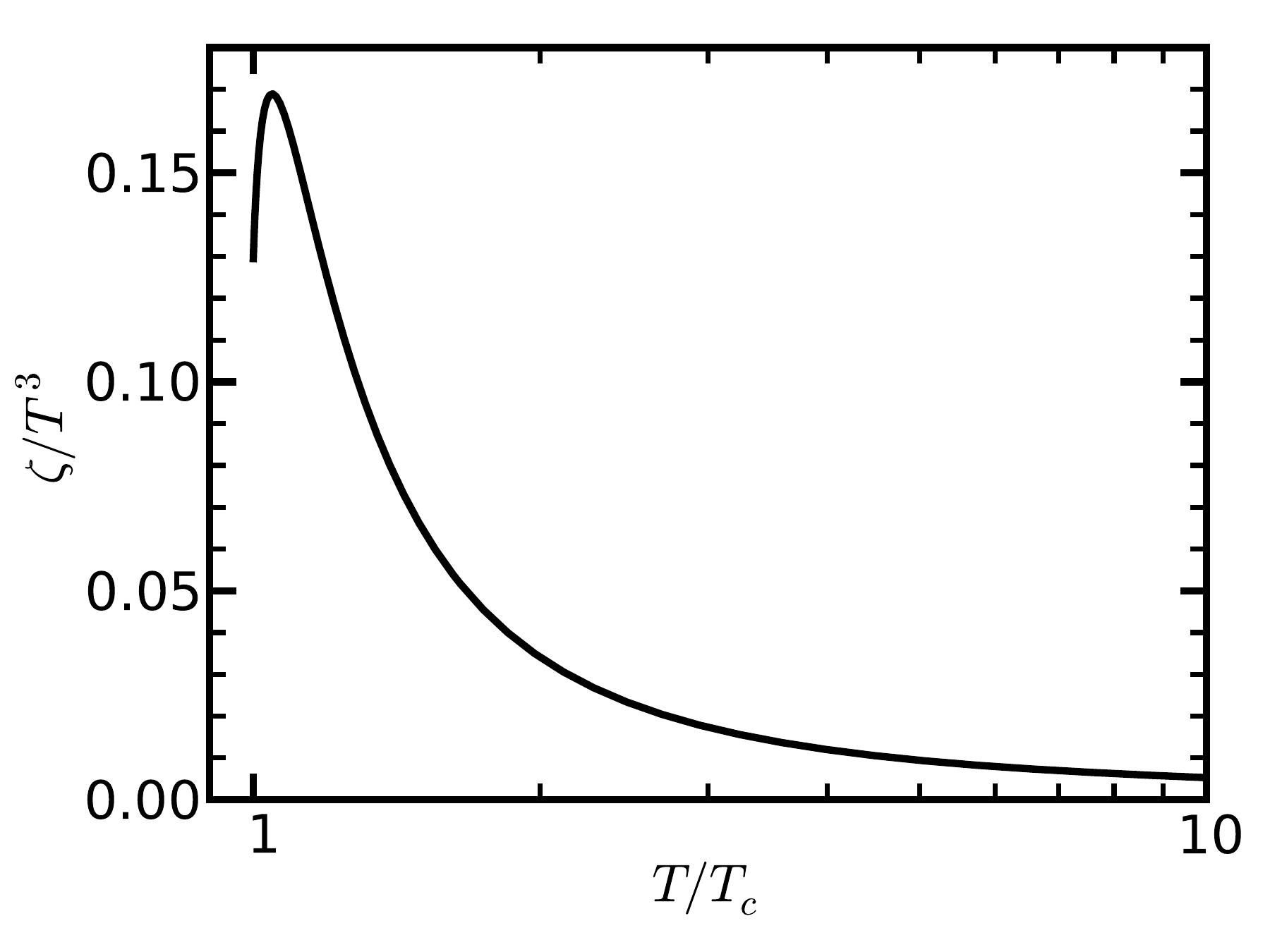} 
\includegraphics[width=0.495\textwidth]{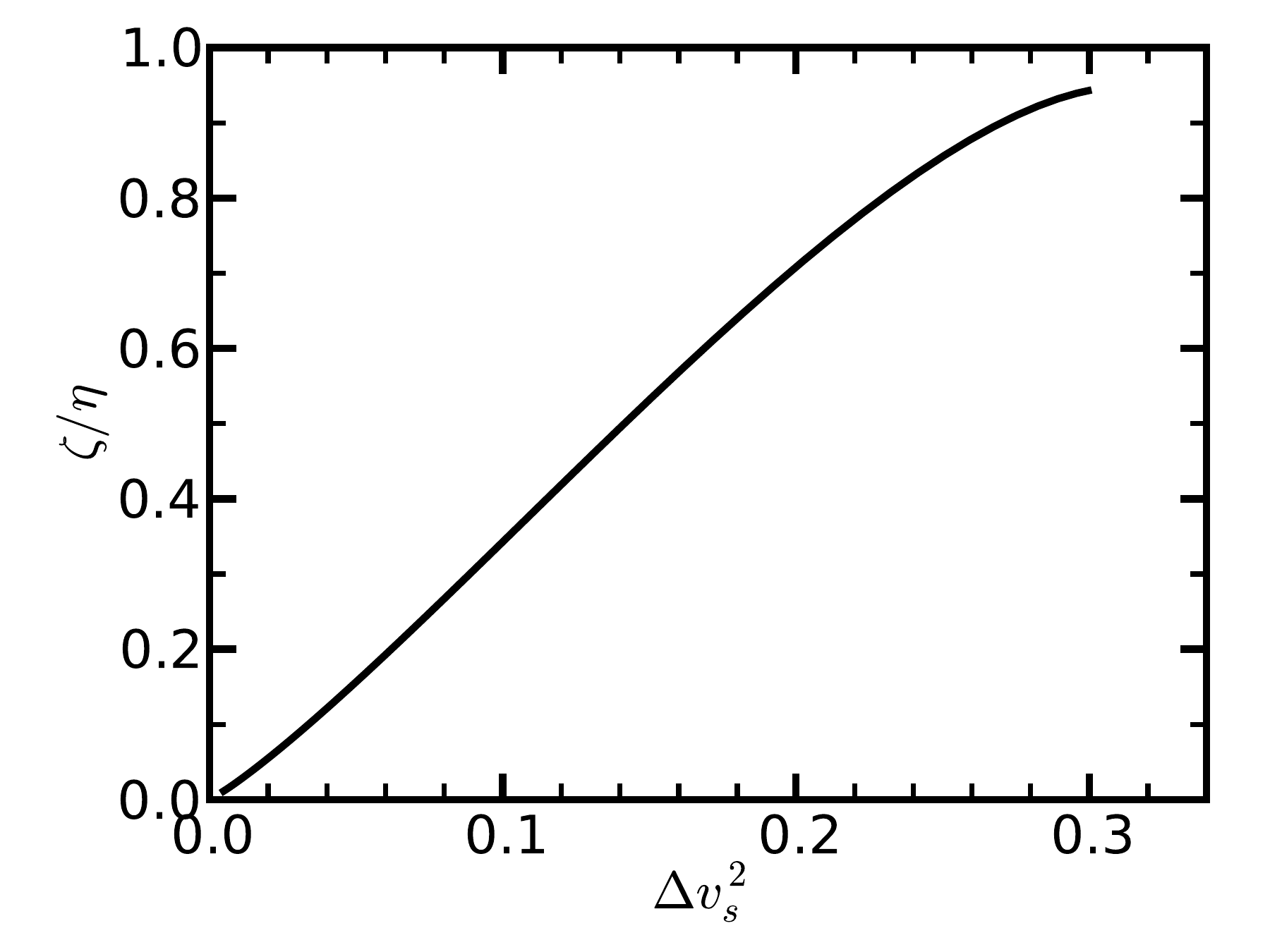}
\end{center}
\caption{The scaled bulk viscosity $\zeta /T^3$
as a function of the temperature (left panel) and the ratio $\zeta / \eta$
as a function of the non-conformality measure (right panel). 
\label{fig.2}}
\end{figure}
Interesting is the relation 
$\zeta / \eta \propto 1.2 \pi \Delta v_s^2$ for $0.025 < \Delta v_s^2 < 0.2$ 
which follows numerically and is specific for the selected potential parameters. 
This corresponds to the temperature interval $1.05 < T/T_c < 2$;
extending the fit to $1.05 < T/T_c < 10$ we find $\zeta / \eta \approx \pi \Delta v_s^2$. 
The IHQCD model \cite{Kiritsis_zeta} yields also $\zeta / \eta \propto 1.2 \pi \Delta v_s^2$,
i.e.\ it is on top of the curve in the right panel Fig.~2, but stops at $\Delta v_s^2(T_c) \approx 0.22$. 

The viscosity ratio accommodates the Buchel bound $\zeta/\eta \geq 2 \Delta v_s^2$ 
\cite{Buchel_bound} and
agrees surprisingly well on a qualitative level with the result of 
\cite{Bluhm} in the interval $1.05 < T/T_c < 2$. 
There, a quasi-particle approach has been employed which needs, beyond the 
equation-of-state adjustment, further input: In \cite{Bluhm} it is the dependence
of the relaxation time on the temperature which causes a change from the linear
relation $\zeta / \eta \propto \Delta v_s^2$ near $T_c^+$, i.e.\ for large 
values of $\Delta v_s^2$, to a quadratic dependence in the weak-coupling regime \cite{Arnold_zeta}
at large temperatures corresponding to small values of $\Delta v_s^2$.
Note also the shift of the linear section of $\zeta / \eta$ in \cite{Bluhm} by a 
somewhat larger off-set which can cause a descent violation of the Buchel bound,
which is not unexpected with respect to \cite{Buchel_bound_violation}.

\section{Robustness of the bulk viscosity}
\subsection{Definition of the transition temperature}
\label{sec:Tc_def}
If one is interested in the thermodynamics of the deconfined phase 
a theoretically sound determination of $T_c$ can be related to the 
Hawking-Page transition and to the construction of \cite{Kiritsis_long}, 
as strictly applied in section \ref{sec:EoS}. Fitting the data \cite{WuppertalBp}, 
we observe \cite{self_v3} $\tilde T_c = (1 + \varepsilon)T_{min}$ with positive 
$\varepsilon < 10^{-2}$ and $\tilde T_c$ from the pressure loop (see Fig.~A.1, inset
in left bottom panel). 
One could be tempted, therefore, to ignore the numerically tiny difference 
of the proper thermodynamic first-order transition temperature 
$\tilde T_c$ and $T_{min}$ and to use $T_{min}$ instead. In fact,
then one can easily reproduce the lattice data \cite{WuppertalBp}, as 
shown in \cite{self_v3} e.g.\ by a potential similar 
to \cite{Gubser_PRL}, distorted by polynomial terms, 
\begin{equation} \label{eq:Gubser_pot}
L^2 V_{IV}(\phi) = -12 \cosh (\gamma \phi) + (6 \gamma^2+\frac12 \Delta [\Delta-4]) \phi^2 
+ \sum_{i = 2}^{5} \frac{c_{2i}}{(2i)!} \phi^{2i},
\end{equation}
whereby the original Gubser-Nellore potential \cite{Gubser}, referred to as $V_I$, 
follows for $c_{2i} = 0$. 

\subsection{Generating nearly equivalent potentials}
\label{sec:equi_pot}
The scheme of employing the holographic principle here consists of mapping 
$V(\phi \in [\phi_0,\phi_H]) \Rightarrow T(\phi_H), s(\phi_H) \Rightarrow s(T)$,\footnote[3]{
Here, the boundary position is denoted by $\phi_0$, being at $\phi = 0$ for the potential 
\eqref{eq:Gubser_pot}, while in the IHQCD model 
\cite{Kiritsis_first1, Kiritsis_first2, Kiritsis_long, Kiritsis_compwithdata} it is at $\phi = -\infty$. 
Because of this, the approximate symmetry of the equation of state under 
constant shifts $\phi \rightarrow \phi + \phi_s$, discussed here, is 
\textit{exact} in IHQCD \cite{Kiritsis}.}
i.e.\ the complete 
non-local potential properties enter the local thermodynamics. Since we are 
interested in $s/T^3$ as a function of $T/T_c$ in the restricted interval 
$T = (1 \ldots 10)T_c$, one can ask whether near-boundary properties of $V(\phi)$ 
are irrelevant. We provide evidence that this is indeed the case, 
at least for $\varepsilon \ll 1$, where one can tentatively 
neglect the difference of $T_{min}$ and $\tilde T_c$, and ignoring the IR behavior. 
To substantiate this claim, let us consider a special one-parameter 
potential $V_s(\phi; \phi_s)$ which contains as relevant part the section 
$V_I(\phi \geq \phi_m)$ where $\phi_m = 0.55$ means a value of $\phi_H$ 
corresponding to $10 \tilde T_c$ determined by the potential $V_I$. 
The relevant section of $V_I$ is now
up or down shifted by a parameter $\phi_s$, and 
$L^2 V_{int}(\phi; \phi_s) = 
-12 + \frac 12 L^2 m_{int}^2(\phi_s)\phi^2 + b(\phi_s) \phi^4$ 
is an interpolating section from the boundary $\phi_0$ to the matching point 
$\phi_m + \phi_s$.  The conditions 
$V_I(\phi_m) = V_{int}(\phi_m + \phi_s; \phi_s)$, 
$V_I^\prime(\phi_m) = V_{int}^\prime(\phi_m + \phi_s; \phi_s)$ 
fix $L^2 m_{int}^2$ and $b$. 
The Breitenlohner-Freedman 
bound $-4 \leq L^2 m_{int}^2 \leq 0$ restricts the possible values of $\phi_s$ for given 
$V_{int}$ and $\phi_m$; 
in our example, 
$-0.165 \leq \phi_s \leq 0.4$. 
To quote a few numbers, 
the left-most shift $\phi_s = -0.165$ yields
$L^2m_{int}^2 = -3.927$, $\Delta = 2.271$, $LT_{min} = 1.81 \times10^{-2}$, while 
the right-most shift $\phi_s = 0.4$ yields
$L^2 m_{int}^2 = -0.098$, $\Delta = 3.975$, $LT_{min} = 3.46 \times10^{19}$.\footnote[4]{
There is a subtlety here: due to the small, but finite, influence of the UV region
$T_{min}^{s} \neq T^{s}(\phi_{H,V_I}^{min} + \phi_s)$, however $\vert T_{min}^{s} - T^{s}(\phi_{H,V_I}^{min} + \phi_s)\vert/T_{min}^{s} < 1.3\times10^{-3}$. 
For the procedure described in the text, 
$s/T^3(T_{min})$ varies between $1.11$ and $1.32$. For $T > T_{min}$, 
(and also if one uses $T_{min}^{s} = T^s(\phi_{H,V_I}^{min} + \phi_s)$)
$T/T_{min}$ and $s/T^3$ stay within the corridors mentioned in the text.}
Despite of a huge variation of $LT_{min}$, 
dimensionless thermodynamic 
quantities $T/T_{min}$ and $s/T^3$ as functions of $\phi_H - \phi_s$ are 
within very narrow corridors with relative variations 
(depending on $\phi_H - \phi_s$ and parametrically on $\phi_s$) of 
less than $4\times 10^{-2}$ for $T/T_{min}$ and $5\times10^{-4}$ for $s/T^3$. 
From the Eling-Oz formula \eqref{eq:zeta_EO}, one infers an analogous 
behavior of $\zeta/\eta$ as a function of $\phi_H - \phi_s$, meaning 
that the potentials $V_s$ deliver a nearly unique equation of state 
and viscosity ratio in the considered temperature interval. 
We therefore argue that all precise fits of $V(\phi)$ 
to lattice data deliver, up to a linear shift, nearly equivalent potentials  
in the selected temperature region and, in particular, 
nearly the same $\zeta/\eta$ vs.\ $\Delta v_s^2$. \\

At the end of this degression on the role of $T_c$ and the conjectured 
robustness of the 
bulk viscosity we mention that we are not able to fit precisely \eqref{eq:v1} 
with parameters \eqref{eq:v1_pars} 
by $V^\prime/V$ emerging from the potential \eqref{eq:Gubser_pot} 
with $\gamma > \sqrt{2/3}$. Apparently, \eqref{eq:Gubser_pot} and the proper 
$T_c$ definition along \cite{Kiritsis_long} with well defined IR behavior 
seem to fail a precise match to the 
data \cite{WuppertalBp}. 
In the Appendix we present a potential which accomodates also lattice data below $\tilde T_c$.

\section{Discussion and Summary}

Inspired by the AdS/CFT correspondence we employ an AdS/QCD hypothesis and
adjust, in a bottom-up approach, the dilaton potential parameters at lattice 
gauge theory thermodynamics data for the pure SU(3) gauge field sector.
We describe several variants to accurately reproduce the
data of \cite{WuppertalBp} in the LHC relevant temperature region 
from $T_c$ up to $10 T_c$. 
Conceptually, the match to the thermal gas solution at $T < T_c$ is most 
satisfactory and can be accomplished by a properly designed dilaton potential, 
which precisely catches the lattice data above $T_c$. Giving up the criteria 
of \cite{Kiritsis_first2, Kiritsis_long} for a zero-temperature confining boundary 
theory with a gapped excitation spectrum in the deep IR, one can 
construct a thermodynamic first-order phase transition with a perfect match 
of lattice data within $(0.7 - 10)T_c$. 
When focusing on $T > T_c$ the Gubser-Nellore potential form 
is comfortable for fitting the lattice data with an 
ad hoc choice of a scale identified with $T_c$. Clearly this latter 
variant ignores the physics of the boundary theory below and at $T_c$. 

Despite of such ambiguities, we find the bulk viscosity at and above $T_c$ as 
fairly robust, with deviations of at most $6\%$ for $T/T_c \leq 1.02$ and otherwise 
less than $2\%$, supposed $T_c$ is a proper 
first-order transition temperature (if not, the bulk viscosity can significantly 
vary, depending upon the choice of the scale, see also \cite{RY_BK}).

Within the non-conformal region $1 \le T/T_c \le 10$, where the non-conformality measure
is $0.2 > \Delta v_s^2 > 0.004$
and the interaction measure is $ 2.48  > I/T^4 > 0.07$,
an almost linear dependence $\zeta / \eta \approx \pi \Delta v_s^2$ 
on the non-conformality measure $\Delta v_s^2$ 
is observed, as already argued in \cite{Buchel} and
found within 
holographic approaches \cite{Buchel_bound, Cherman_Yarom} and in \cite{Bluhm} 
within a quasi-particle approach to the pure gauge sector of QCD. 

We mention further that one can identify a relevant 
section of the potential which determines the equation of state in a 
selected temperature interval. Shifting, within certain limits that relevant 
section, the equation of state and the bulk viscosity are marginally 
modified. 


Extensions towards including quark degrees of freedom and subsequently 
non-zero baryon density, i.e.\ to address full QCD, have been outlined 
and explored in \cite{Gubser_quarks}. The Veneziano limit of QCD 
is investigated in a more string theory inspired setting in \cite{Kiritsis_VQCD}. 
Incorporating additional degrees of freedom (which are aimed at mimicking
an equal number of quarks and anti-quarks) within the present set-up,
one essentially has to lower $G_5 / L^3$ in adjusting the extensive and intensive
densities. Since the viscosities scale with $L^3 /G_5$ \cite{Gubser_visc}
(as the entropy density does, too)
the corresponding ratios $\zeta/s$ and $\zeta/\eta$ would stay unchanged,
if the same potential would apply and the same behavior of the sound
velocity would be used as input.
However, as stressed above, $\zeta / \eta$ depends rather sensitively 
on the actual potential $V(\phi)$ and its parameters. 
Since QCD does not display a first-order phase transition 
at zero baryon density, 
dedicated separate investigations are required to adjust the dilaton potential 
to current lattice data. 
(The results in \cite{Gubser_visc} yield 
$\zeta/\eta \approx 0.98 \pi \Delta v_s^2$ for $\Delta v_s^2 < 0.28$ 
with a maximum of $\zeta/\eta \approx 0.75$ at $\Delta v_s^2 \approx 0.26$, i.e.\ 
values comparable to the pure glue case.) 

On the gravity side, inclusion of
terms beyond the Hilbert action would cause a temperature dependence of the ratio
$\eta / s$ \cite{Durham} 
which is needed to furnish the transition into the weak-coupling
regime \cite{eta_perturbative} at large temperatures. 
It is an open question whether such higher-order curvature corrections also 
lead to a quadratic dependence of the viscosity ratio on the non-conformality 
measure \cite{Arnold_zeta}.

In summary, we adjust the dilaton potential exclusively at new lattice data
for SU(3) gauge theory thermodynamics
and calculate holographically the bulk viscosity. The ratio of the bulk to shear 
viscosity obeys, in the strong-coupling regime, 
a linear dependence on the non-conformality measure
for temperatures above $1.05 T_c$, while at $T_c$ it has a maximum
of $0.94$. Our result, which is based on some fine tuning of the dilaton 
potential to precision lattice data, agrees well with previous holographic 
approaches based on former lattice data, such as the IHQCD model, or studies with 
the Gubser-Nellore potential types which envisaged qualitatively 
capturing QCD features. 

It would be interesting to employ the numerical findings 
of our holographically motivated guess, even if they are related to the pure 
gauge theory (with the disclaimers mentioned in the introduction), e.g.\ 
in the modellings \cite{DS, DS_prime, Dima} of heavy-ion collisions to 
elucidate their impact on observables. 
Our potential(s) may also serve as a suitable background e.g.\ for various
holographic mesons.

Acknowledgements: Inspiring discussions with M. Bluhm, M. Huang, E. Kiritsis, 
K. Redlich, C. Sasaki, 
U. A. Wiedemann, P. M. Heller and L. G. Yaffe are gratefully acknowledged. 
The work is supported by BMBF grant 05P12CRGH1 and European Network HP3-PR1-TURHIC.

\setcounter{footnote}{4}
\setcounter{figure}{0}
\begin{appendix}

\section{Including confined-phase lattice data}
\label{sec:appendix}
The potential $v_1$ in \eqref{eq:v1} can be modified
to reproduce also the presently available lattice data in the confined phase:
\begin{equation} \label{eq:v2}
 v_2 = 
 \begin{cases}
   \frac{- L^2 M^2}{12}\phi + i_1 \phi^3 &\text{ for } \phi \leq \phi_m,  \\
   \gamma + s_1[\tanh(s_1 (\phi - s_2)) - 1] + p_1 e^{p_2(\phi - p_3)^2} &\text{ for } \phi \geq \phi_m.
 \end{cases}
\end{equation}
This parametrization is inspired 
by the desired behavior of $v_s^2$ as function of $\phi_H$.\footnote{
The parametrization \eqref{eq:v2} is superior to the one 
given in the appendix of \cite{self_v3}, since a better 
description of $v_s^2$ for $T < T_c$ is accomplished.} 
Performing a fit to lattice data for $0.7 \leq T/T_c \leq 10$ and identifying $T_c$
with $\tilde T_c$, which is determined by the intersection of the high-temperature and low-temperature 
branches of the pressure \eqref{eq:p} combined with $T(\phi_H)$,
we find the parameters 
\begin{equation} \label{eq:v2_pars}
\begin{tabular}{c|c|c|c|c|c|c|c|c}
   $v$      & $\phi_{m}$ &$s_{1}$  &$s_{2}$ & $\gamma$   & $p_{1}$& $p_{2}$& $p_{3}$ & $G_5/L^3$ \\ \hline
  $v_{2a}$  & 2.3523     & 0.4452  & 6.9382 &$\sqrt{2/3}$& 0.7526 & 0.1707 & 4.6707  & 1.1125 \\ \hline
  $v_{2b}$  & 2.3171     & 0.4259  & 6.5929 & 0.7979     & 0.6982 & 0.1864 & 4.6011  & 1.1116 
\end{tabular} .
\end{equation}
The resulting equation of state is exhibited in Fig.~A.1.
\begin{figure}[!h]
\begin{center}
\includegraphics[width=0.495\textwidth]{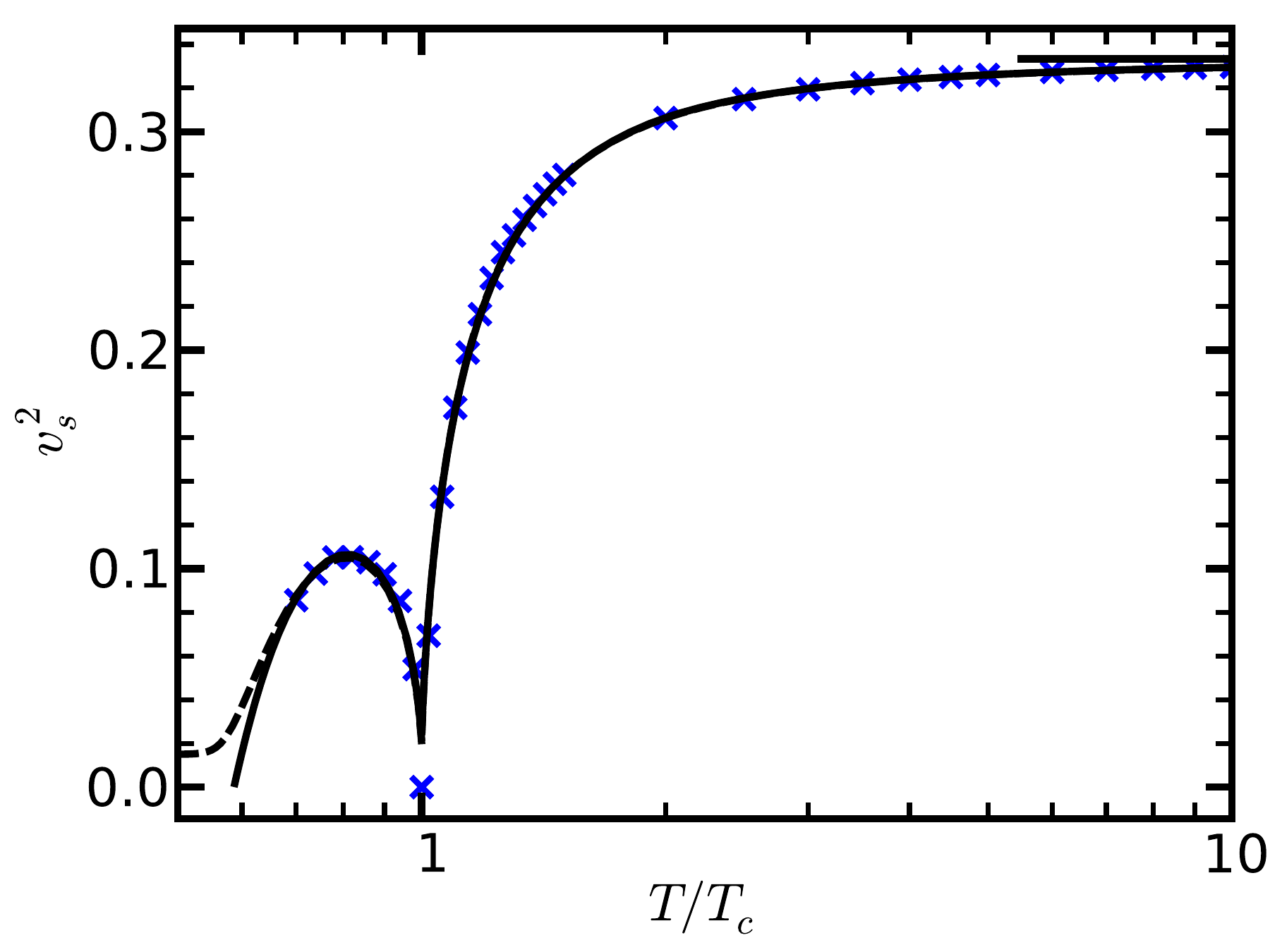}
\includegraphics[width=0.495\textwidth]{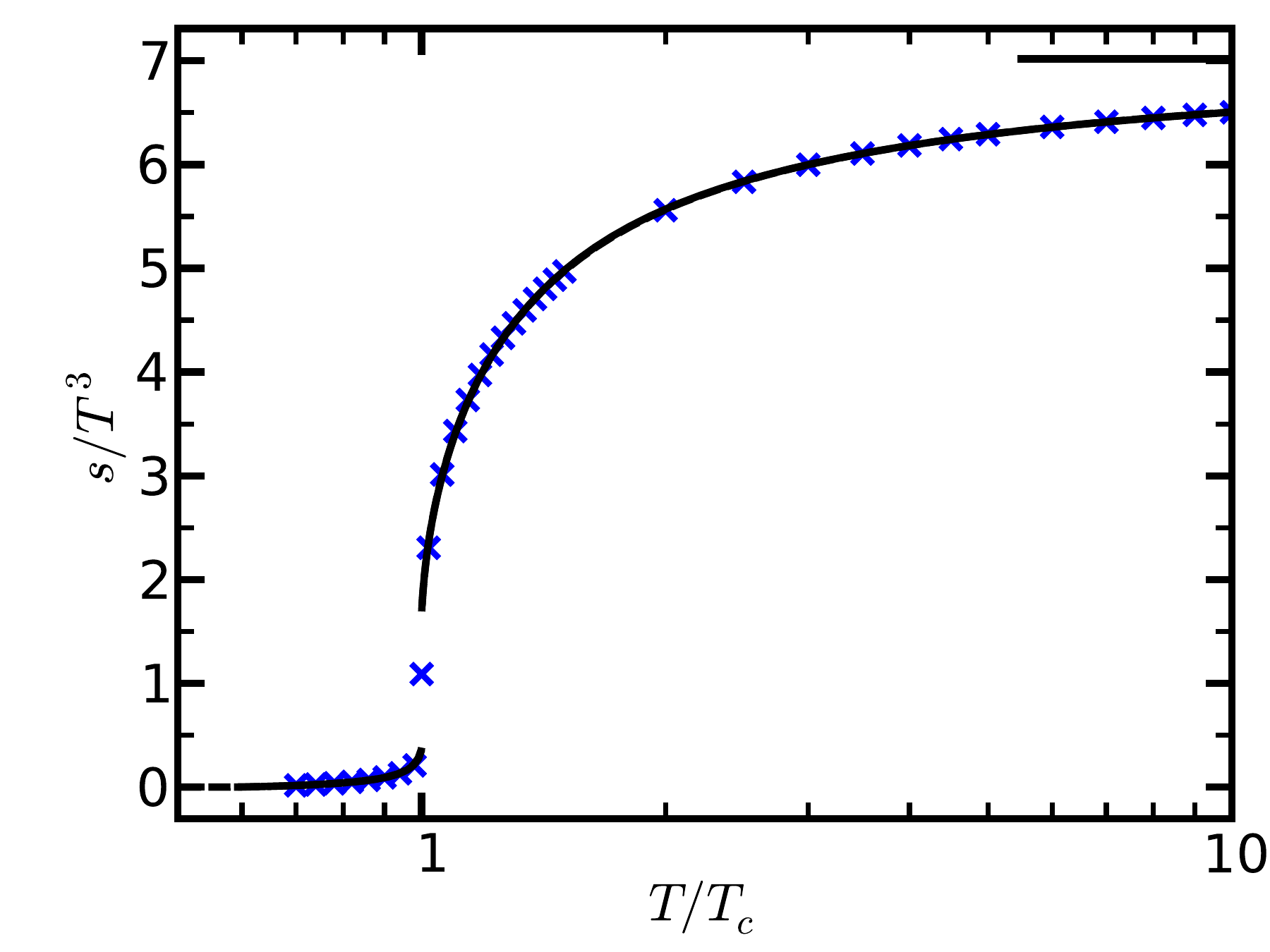}
\includegraphics[width=0.495\textwidth]{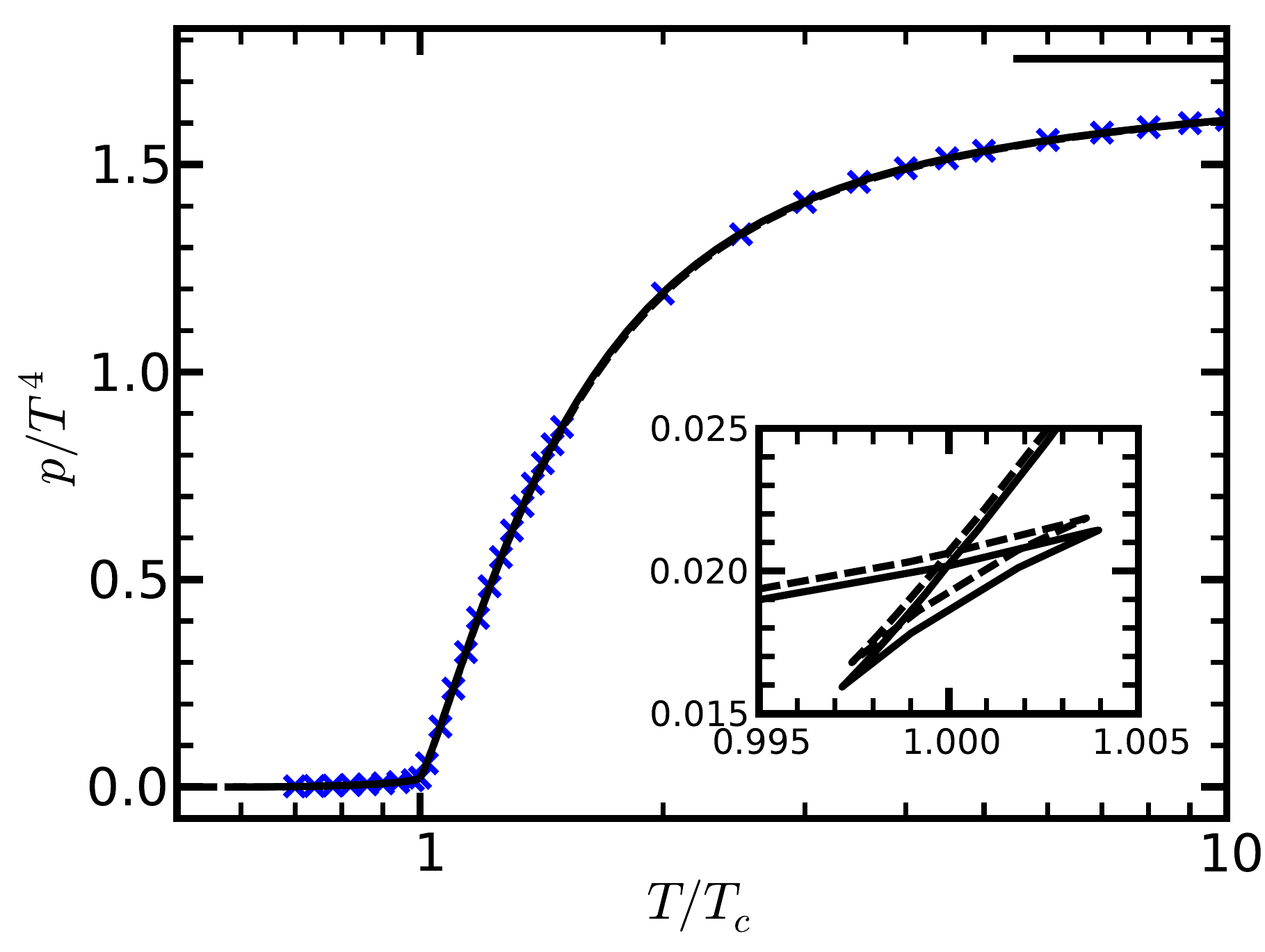}
\includegraphics[width=0.495\textwidth]{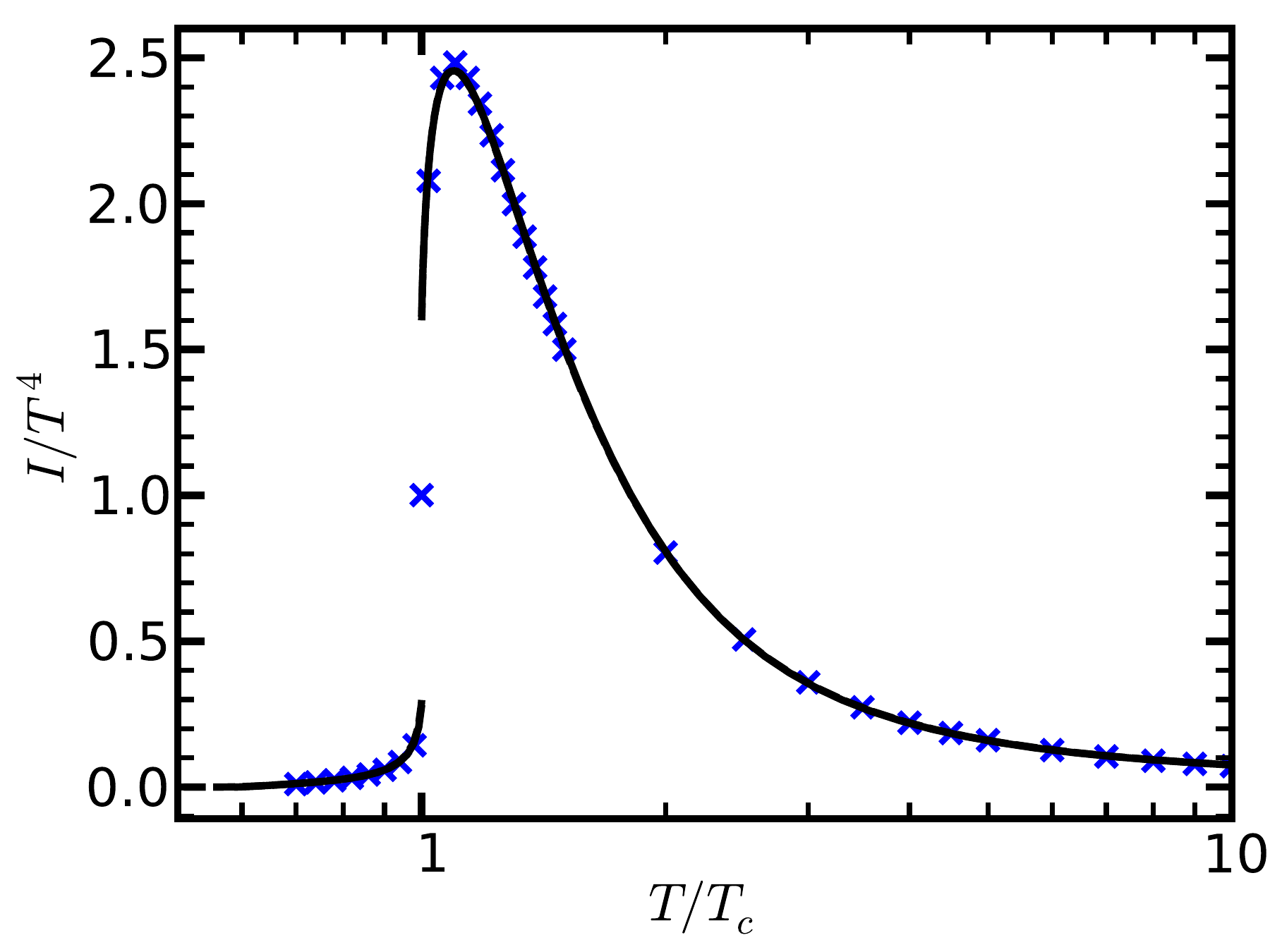}
\end{center}
\caption{
The same as Fig.\ 1 but extending the lattice data points and model calculation 
into the confined phase 
for the potential $v_2$ \eqref{eq:v2} with 
parameters \eqref{eq:v2_pars}. 
Solid curves: $v_{2a}$, dashed curves: $v_{2b}$. 
The un-/metastable branches are not plotted, unless in the inset of the pressure 
panel, where the standard loop structure is displayed for $p/T^4$ with $p$ 
calculated from \eqref{eq:p}.
\label{fig.3}}
\end{figure}
In the direct vicinity of $T_c$ the model calculation deviates 
from the lattice data on a $5\%$ level in the high- and 
low-temperature phases; otherwise the fit is near-perfect. 
Unlike the potential $v_1$ \eqref{eq:v1} which faciliates 
a monotonous increase of $LT(\phi_H)$ for $\phi_H > \phi_H^{min}$, 
$LT(\phi_H)$ from $v_{2a}$ (with fixed $\gamma = \sqrt{2/3}$) runs to a constant value, while 
for $v_{2b}$ it is dropping, see Fig.\ A.2. That is the potentials 
$v_{2a}$ and $v_{2b}$ leave the IR physics of the boundary 
theory unsettled, which however does not play any role 
for the description of the lattice data for $T > 0.7T_c$ as 
seen from Fig.\ A.1. 
The potential $v_{2a}$ can be regarded as the best compromise between 
two mutually exclusive options: 
$v_1$, a zero-temperature confining and gapped boundary theory and 
$v_{2b}$, a boundary theory with smooth and finite pressure for $0 < T < T_c$;
in the classification of \cite{Kiritsis_long}, the model $v_{2a}$ is 
zero-temperature confining and has a partially discrete spectrum.
For both parameter sets \eqref{eq:v2_pars} we find the scaled latent heat $\Delta s(T_c)/T_c^3 \approx 1.3$  
which compares well with $\Delta s(T_c)/T_c^3 \approx 1.4$ found in lattice calculations
(see \cite{WuppertalBp} and references therein). 

The bulk viscosity resulting from the ansatz \eqref{eq:v2} with the 
parameters \eqref{eq:v2_pars} is exhibited in Fig.~A.2. 
The maximum $\zeta/\eta \approx 1$ lies at $\Delta v_s^2 \approx 0.31$. 
We notice the jump at $T_c$ due to the first-order phase transition; $\zeta/T^3$ 
is rapidly dropping for smaller temperatures; $\zeta/\eta$ vs.\ $\Delta v_s^2$ 
displays a hook which we would not consider a reliable result since 
the setting at $T < T_c$ might not be trustworthy. 
Below $T_c$, 
in the interval $0.76 \lesssim T/T_c \lesssim 0.998$, 
the viscosity ratio $\zeta/\eta$ violates the 
Buchel bound (see right panel of Fig.~A.2). A similar behavior 
was found in \cite{Gubser_visc} for the potential $V_I$ adjusted to 
the equation of state of 2+1 flavor QCD.

\begin{figure}[!h]
\begin{center} 
\includegraphics[width=0.495\textwidth]{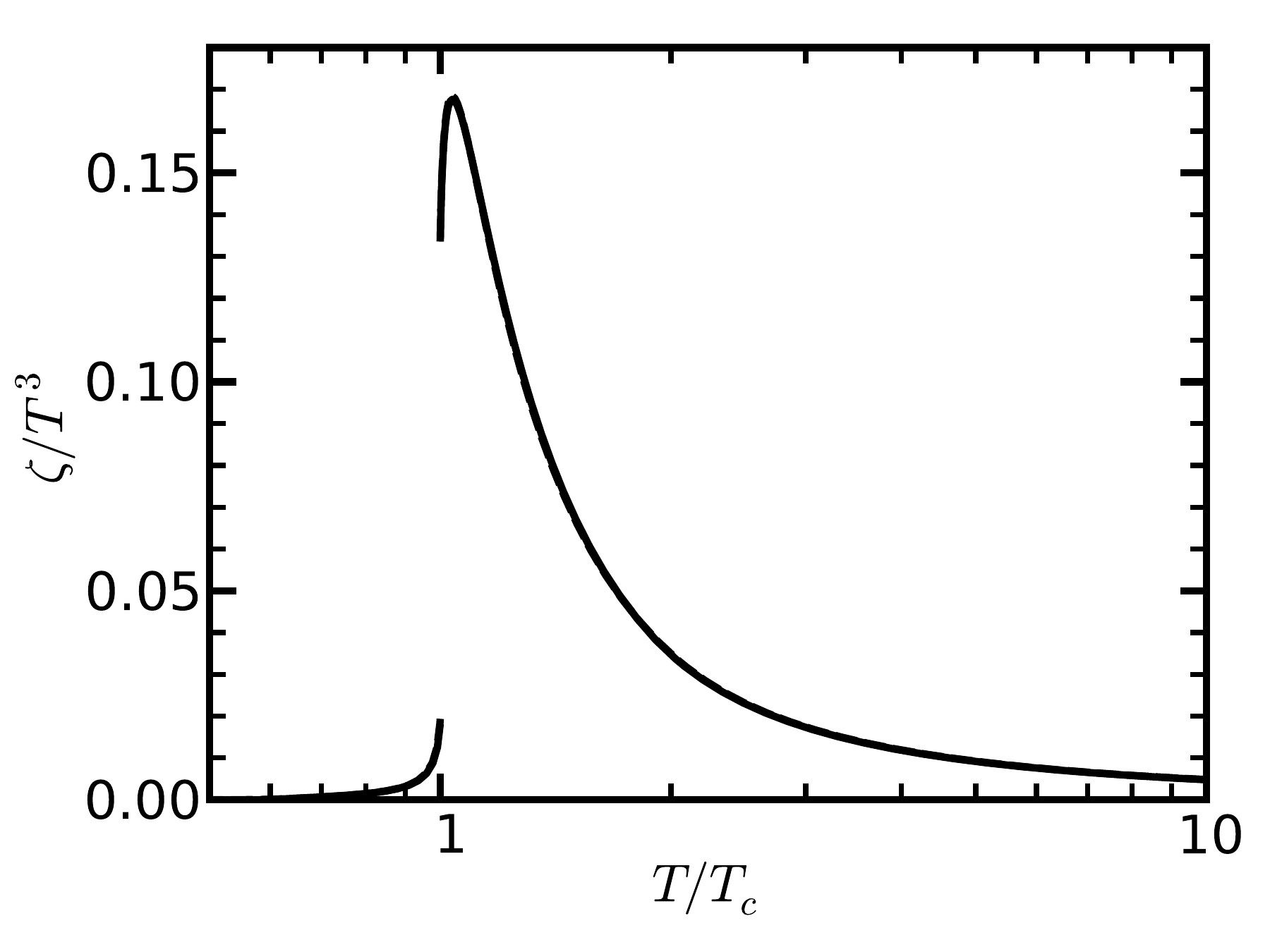} 
\includegraphics[width=0.495\textwidth]{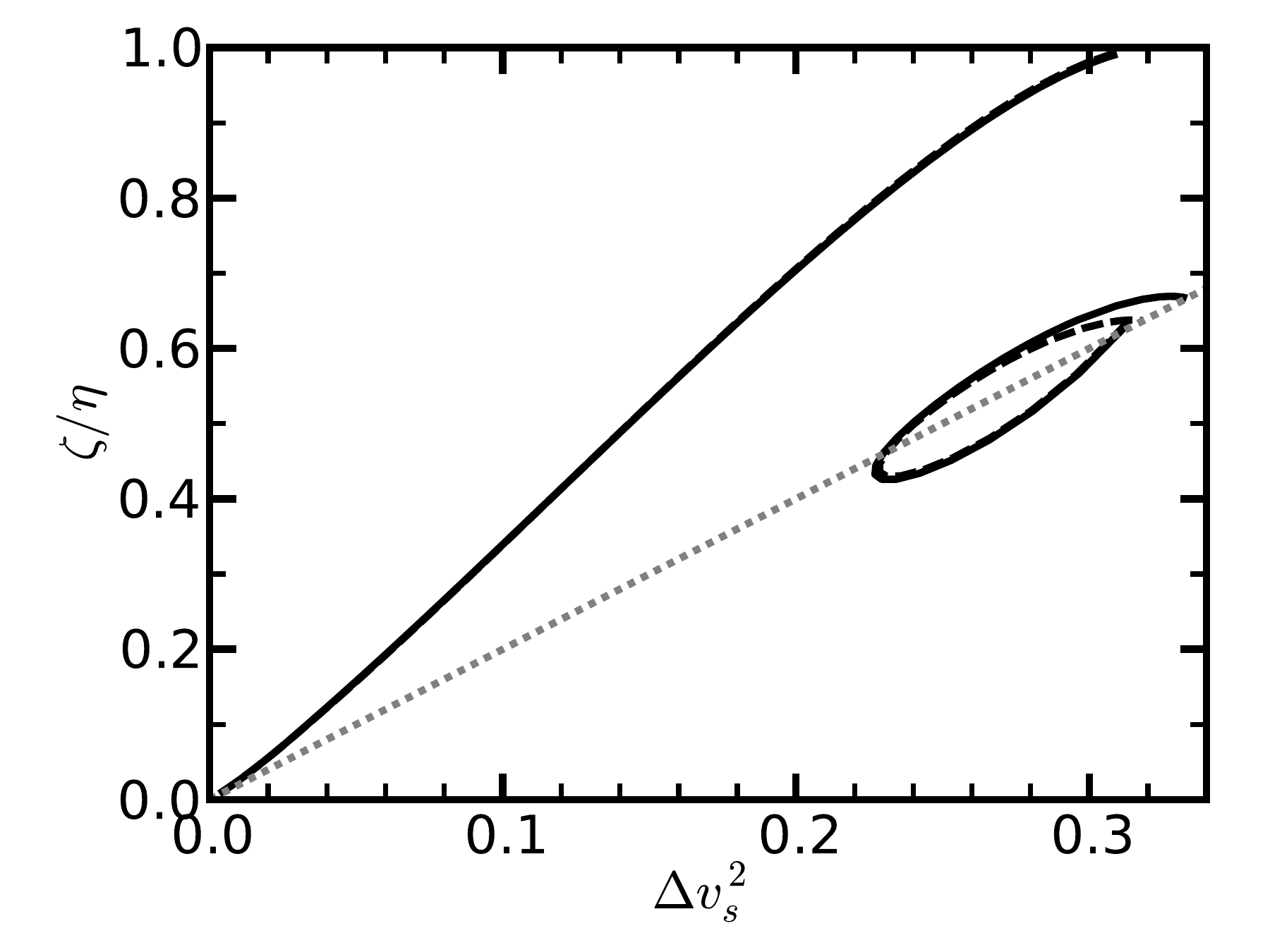}
\end{center}
\caption{The scaled bulk viscosity $\zeta /T^3$
as a function of the temperature (left panel) and the ratio $\zeta / \eta$
as a function of the non-conformality measure (right panel). 
Line codes are the same as in Fig.\ A.1. 
The grey dotted line in the right panel depicts the Buchel bound. 
\label{fig.4}}
\end{figure}

Figure A.3 summarizes the dependence of the temperature as a function of $\phi_H$. 
The global minimum for $v_1$ \eqref{eq:v1} is quite shallow (the anticipated U shape 
becomes better evident when displaying $LT$ as a function of $\log \phi_H/\phi_H^{min}$). 
The local minima for $v_{2a,b}$ \eqref{eq:v2} are also very shallow. 
Thus, $T_{min} \approx T_c$ or $\tilde T_c$ follows.

\begin{figure}[!h]
\begin{center}
\includegraphics[width=0.95\textwidth]{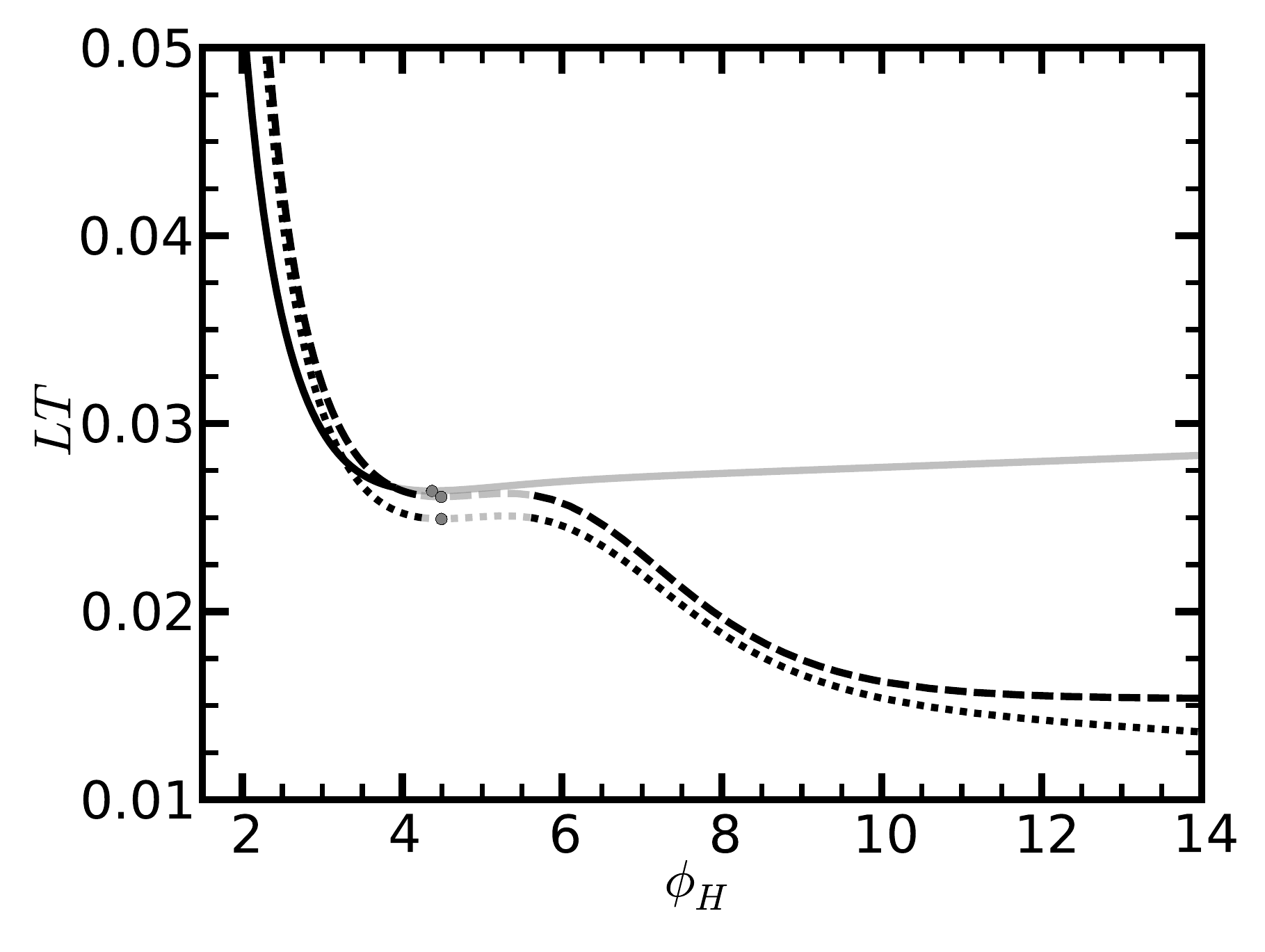}
\end{center}
\caption{The temperature as a function of $\phi_H$
for the potentials $v_1$ \eqref{eq:v1} with parameters \eqref{eq:v1_pars} (solid curve) 
as well as $v_{2a}$ (dashed curve) and $v_{2b}$ (dotted curve), 
see \eqref{eq:v2} and \eqref{eq:v2_pars}. 
Light grey portions of the curves denote the un-/metastable regions of the 
equation of state, while dots mark the positions of $\phi_H^{min}$.
\label{fig.5}}
\end{figure}
\end{appendix}

\pagebreak[4]

\end{document}